\documentclass[12pt]{article}
\usepackage[dvips]{graphicx}
\usepackage{amssymb}
\usepackage{amsmath}
\usepackage{here}
\usepackage{color}
\usepackage{subfigure}
\usepackage{cite}

\newcommand{\bear}{\begin{array}}  
\newcommand {\eear}{\end{array}}
\newcommand{\bea}{\begin{eqnarray}}   
\newcommand{\eea}{\end{eqnarray}}
\newcommand{\beq}{\begin{equation}}   
\newcommand{\eeq}{\end{equation}}
\newcommand{\bef}{\begin{figure}}  \newcommand 
{\eef}{\end{figure}}
\newcommand{\bec}{\begin{center}}  \newcommand 
{\eec}{\end{center}}

\begin{document}

\vskip 2.0cm

\begin{center}

{\large \bf 
Anisotropies in the gravitational wave background as a probe of the cosmic string network \\
}

\vskip 1.2cm

Sachiko Kuroyanagi$^{a,b}$
Keitaro Takahashi$^{c}$
Naoyuki Yonemaru$^{c}$
Hiroki Kumamoto$^{c}$

\vskip 0.4cm

{\it $^a$Department of Physics, Nagoya University, Chikusa, Nagoya 464-8602, Japan}\\
{\it $^b$Institute for Advanced Research, Nagoya University, Chikusa, Nagoya 464-8602, Japan}\\
{\it $^c$Faculty of Science, Kumamoto University, 2-39-1 Kurokami,
Kumamoto 860-8555, Japan}\\
\date{today}

\end{center}

\vskip 0.2cm

\begin{abstract}
  Pulsar timing arrays are powerful tools to test the existence of
  cosmic strings by searching for the gravitational wave (GW)
  background.  The amplitude of the background connects to information
  on cosmic strings such as the tension and string network properties.
  In addition, one may be able to extract more information on the
  properties of cosmic strings by measuring anisotropies in the GW
  background.  In this paper, we provide estimates of the level of
  anisotropy expected in the GW background generated by cusps on
  cosmic strings.  We find that the anisotropy level strongly depends
  on the initial loop size $\alpha$, and thus we may be able to put
  constraints on $\alpha$ by measuring the anisotropy of the GW
  background.  We also find that certain regions of the parameter
  space can be probed by shifting the observation frequency of GWs.
\end{abstract}

\section{Introduction}
Cosmic strings are one-dimensional topological defects, which arise
naturally in field theories \cite{Kibble:1976sj,Vilenkin}, as well as
in scenarios of the early Universe based on superstring theory
\cite{Sarangi:2002yt,Jones:2003da,Dvali:2003zj}. One promising
strategy to test for their existence is to search for gravitational
wave (GW) emission from them. In particular, strong GW bursts are
emitted from nonsmooth structures such as cusps and kinks
\cite{Damour:2000wa} and overlapped bursts form a stochastic GW
background over a wide range of frequencies
\cite{Damour:2001bk,Damour:2004kw,Siemens:2006yp,DePies:2007bm,Olmez:2010bi,Sanidas:2012ee,Sanidas:2012tf,Binetruy:2012ze,Kuroyanagi:2012wm,Kuroyanagi:2012jf,Sousa:2013aaa}.

Pulsar timing arrays uniquely probe the GW background at nanohertz
frequencies
\cite{Arzoumanian:2015liz,Shannon:2015ect,Lentati:2015qwp,Verbiest:2016vem}. GWs
affect the times of arrival (ToAs) of pulses so that the residuals of
the ToAs indicate the existence of GWs.  In the case of the stochastic
GW background, cross-correlations of the residuals between multiple
pulsars are taken to reduce the noise, and the correlation coefficient
as a function of the angle between two pulsars is called the Hellings
and Downs curve \cite{Hellings:1983fr}. The current limits on the
strain amplitude of the GW background have already produced strong
constraints on cosmic strings
\cite{Arzoumanian:2015liz,Lentati:2015qwp}. In the future, the
International Pulsar Timing Array \cite{Hobbs:2009yy} and the Square
Kilometre Array (SKA) \cite{Janssen:2014dka} will enhance the
sensitivity and offer the best opportunity to search for cosmic
strings.

Recently, a method to analyze anisotropies in the GW background has
been progressively developed
\cite{Mingarelli:2013dsa,Cornish:2013aba,Cornish:2014rva}. The
anisotropies can arise due to the finiteness of the GW sources and
reflect the number of sources and their distribution. In the presence
of anisotropies, the cross-correlation of the timing residuals between
different pulsars deviates from that of the Hellings and Downs curve,
which is derived assuming an isotropic GW background.  Simulation
studies have showed that we would obtain substantial evidence for the
anisotropy signal when the signal-to-noise ratio is higher than $10$
\cite{Taylor:2013esa}.  The European Pulsar Timing Array
\cite{Taylor:2015udp} reported that the spherical harmonics multipole
component of the GW amplitude for $\ell>0$ is less than $40\%$ of the
isotropic component with $95\%$ confidence.  Although this analysis is
developed in the context of a GW background from supermassive black
hole (SMBH) binaries, we expect to apply it to the background from cosmic
strings as well. Information on the source population would help us to
understand the string network evolution.

In this paper, we perform theoretical estimates of the expected level
of anisotropy in a GW background composed of a superposition of GW bursts
originating from cusps on string loops, which are typically the
dominant source of the GW background at nanohertz frequencies.
\footnote{ The dominant source could be taken over by kinks on
  infinite strings \cite{Kawasaki:2010yi} or loops
  \cite{Binetruy:2010cc} depending on the string network properties,
  which is beyond the scope of our paper. }  First, we calculate the
number density of cosmic string loops as a function of the redshift
using the velocity-dependent one-scale model
\cite{Martins:1996jp,Martins:2000cs} and convert it to the rate of the
GW bursts coming to the Earth.  Then we generate data sets of GW
backgrounds by randomly distributing the burst events in the sky, and
calculate the anisotropy level of the GW background using the formalism
established in Ref. \cite{Taylor:2013esa}.

This paper is organized as follows: In Sec. \ref{sec:formulation}, we
describe the one-scale model to obtain redshift distributions of the
loop-number density.  Then we present a model to convert the number
density to the rate of GW emissions, which we use to construct the
data sets of GW backgrounds.  In Sec. \ref{sec:anisotropy}, we briefly
present the formalism to estimate a level of anisotropy by decomposing
the angular distribution of the GW power on the sky into multipole
moments.  Then we present the results with the dependence of the initial
loop size $\alpha$, which is the key parameter to produce a large
anisotropy.  Section \ref{sec:conclusion} is devoted to conclusions.

\section{GW background from cusps on cosmic string loops}
\label{sec:formulation}

The basic components of a cosmic string network are loops and infinite
strings.  Loops are continually formed by the intersection of cosmic
strings, and the typical loop size at formation is often characterized
as $\sim\alpha H(t)^{-1}$, where $H(t)$ is the Hubble scale at loop
formation.  Estimates in earlier works have suggested that the
initial loop size is determined by gravitational backreaction, and
$\alpha$ has values smaller than $\sim G\mu$, where $G\mu$ is the
tension of cosmic strings \cite{Bennett:1987vf,Allen:1990tv,Vincent:1996rb,Siemens:2002dj},
while recent simulations suggest that a significant fraction of loops
are produced at scales roughly a few orders of magnitude below the
horizon size $\alpha\sim O(0.1)$ \cite{Vanchurin:2005pa,Ringeval:2005kr,Martins:2005es,Olum:2006ix}.
Since the loop size distribution is still an ongoing topic \cite{Polchinski:2006ee,Dubath:2007mf,Vanchurin:2007ee,Lorenz:2010sm,BlancoPillado:2011dq,Blanco-Pillado:2013qja},
we take $\alpha$ as a free parameter, and in fact $\alpha$ is a key
parameter for the level of anisotropy.

The other parameters, such as the tension $G\mu$ and reconnection
probability $p$, are also important for the evolution of the cosmic
string network.  The value of the string tension $G\mu$ depends
strongly on the generation mechanism.  For field-theoretic cosmic
strings, $\mu$ is roughly the square of the energy scale of the phase
transition which produces cosmic strings.  For cosmic superstrings,
$\mu$ is determined by the fundamental string scale as well as the
warp factor of the extra dimension, and it can take a broad range of
values.  The tension determines the energy loss of loops through the
emission of GWs, and it relates to the lifetime of loops.  The value
of reconnection probability $p$ also depends on the origin of cosmic
strings.  It is essentially $1$ for field-theoretic cosmic strings,
while it can be smaller than $1$ in the case of cosmic superstrings
because of the effect of extra dimensions \cite{Jackson:2004zg,Hanany:2005bc,Jackson:2007hn}.  Analysis in
Ref. \cite{Jackson:2004zg} suggests the value is in the range of
$0.1\leq p \leq 1$ for D-strings and $10^{-3} \leq p \leq 1$ for
F-strings. The reduced reconnection probability decreases the loss of
infinite string length into loops and eventually enhances the density
of the string network \cite{Sakellariadou:2004wq,Avgoustidis:2005nv}.

It has been shown that cusps on string loops generically arise once
per oscillation time \cite{Kibble:1982cb} and emit strong GW bursts
\cite{Damour:2000wa}.  The typical frequency of GWs is determined by the
loop size at the emission, and so it does depend on $\alpha$.  Overlapped
GW bursts are detectable as a GW background at nanohertz frequencies
when $\alpha$ is not too small \cite{DePies:2007bm}.  In order to
predict the amplitude and anisotropy level of the GW background, we
need to estimate the number and amplitude of GW bursts coming to the
Earth during the observation period of the pulsar timing arrays.
Here, we describe a theoretical model which is used to obtain the
number density of loops and to convert it to the GW rate.

\subsection{Cosmic string network}
Our calculation of the string network evolution is based on the
velocity-dependent one-scale model
\cite{Martins:1996jp,Martins:2000cs}.  In this model, the string
network of infinite strings is characterized by a correlation length
$\xi$, which corresponds to the typical curvature radius and interval
of infinite strings.  Then the total length $L$ of infinite strings in
volume $V$ is given by $L=V/\xi^2$, and the average string energy
density is given by $\rho=\mu/\xi^2$.  From the equation of energy
conservation, one can obtain an evolution equation for $\rho$, whereas
the equations of motion for the Nambu-Goto string yield a equation for
the evolution of the typical root mean square velocity $v$ of infinite
strings.  By defining $\gamma\equiv\xi/t$, the resulting equations are
\beq
\frac{t}{\gamma}\frac{d\gamma}{dt}=-1+\nu +\frac{\tilde{c}pv}{2\gamma}+\nu v^2,
\label{gammaeq}
\eeq 
\beq
\frac{dv}{dt}= (1-v^2)H (\frac{k (v)}{\nu\gamma}-2v),
\label{veq}
\eeq
where $k(v)=\frac{2\sqrt{2}}{\pi}\frac{1-8v^6}{1+8v^6}$, $H\equiv
\dot{a}/a$, and the scale factor $a$ is parametrized as $a (t)\propto
t^{\nu}$.  The third term on the right-hand side of
Eq. (\ref{gammaeq}) represents the loss of energy from infinite
strings by the production of loops.  The constant parameter
$\tilde{c}$ represents the efficiency of loop formation and is set to
be $\tilde{c}=0.23$, and the effect of the reconnection probability
$p$, which deviates from $1$ for cosmic superstrings, can be simply
included by replacing $\tilde{c}$ with $\tilde{c} p$.

Cosmic string networks are known to evolve towards a so-called scaling
regime in which the characteristic length of infinite strings $\xi$
evolves at a rate proportional to the Hubble scale and the number of
them in a Hubble horizon remains constant.  The above sets of
equations indeed have asymptotic solutions, which can be obtained by
setting $d\gamma/dt$ and $dv/dt$ to be $0$.  For example, for $p=1$,
we obtain $\gamma_r=0.27$ and $\gamma_m=0.62$, where $\gamma_r$ and
$\gamma_m$ are the values in the radiation- and matter-dominated eras,
respectively.  Since the effect of the time dependence of $\gamma$
around the matter-radiation equality is small for the settings in this
paper, we approximate $\gamma$ as a step function,
\beq 
\gamma (z)=
\begin{cases}
\gamma_r & ;z>z_{\rm eq} \\
\gamma_m & ;z<z_{\rm eq}
\end{cases}
,
\label{gamma_rm}
\eeq
where $z=1/a(t)-1$ represents the redshift and $z_{\rm eq}$ is the redshift at the matter-radiation equality.

In the scaling regime, infinite strings continuously lose their length
by formation of loops, and the length to lose in a Hubble volume
per Hubble time is comparable to the length of infinite strings in a
Hubble volume.  Assuming that the size of the loops formed at time $t$ is
given by $\alpha t$, the number density of loops generated between
time $t$ and $t+dt$ is
\beq
\frac{dn}{dt}(t)dt=\frac{dt}{\alpha \gamma^2 t^4}.
\label{dn_dt}
\eeq
By taking into account the dilution of the number density due to
cosmic expansion $\propto a^{-3}$, the number density of loops formed
between $t_i$ and $t_i+dt_i$ at time $t$ is given by
\beq
\frac{dn}{dt_i} (t,t_i)dt_i=\frac{dt_i}{\alpha \gamma^2 t_i^4}\left (\frac{a (t_i)}{a (t)}\right)^3.
\label{dn_dt2}
\eeq

\subsection{Rate of GW bursts from cusps on string loops}
After loop formation, the loop continues to shrink by emitting energy
as GWs and eventually vanishes.  The length of a loop at time $t$
formed at $t_i$ is written as
\beq
l (t,t_i)=\alpha t_i - \Gamma G\mu  (t-t_i), \label{length}
\eeq
where $\Gamma$ is a constant which represents the efficiency of GW
emission and we take $\Gamma=50$.
The Fourier amplitude of a GW burst $\tilde{h}(f)=\int ~ dt e^{2\pi ift}h(t)$ from a cusp is formulated in Refs.
\cite{Damour:2000wa,Damour:2001bk,Damour:2004kw} and given by
\footnote{Note that Refs.
\cite{Damour:2000wa,Damour:2001bk,Damour:2004kw} use the logarithmic Fourier transform, and the equation has a difference of factor $f^{-1}$ in our definition $\tilde{h}(f)$. }
\beq
\tilde{h} (f,z,l)\simeq \frac{G\mu l}{ ( (1+z)fl)^{1/3}r (z)f}, \label{strain}
\eeq
where $r (z)=\int^z_0 dz^{\prime}/H (z^{\prime})$.  
Using the above two equations, the loop length $l$ and the generation
time $t_i$ can be given as functions of $\tilde{h}$, $z$, and $f$:
\beq
t_i (f,\tilde{h},z)=\frac{l (f,\tilde{h},z)+\Gamma G\mu t (z)}{\alpha+\Gamma G\mu}, 
\label{t_i}
\eeq
\beq
l (f,\tilde{h},z)=\left (\frac{\tilde{h}r (z)}{G\mu} (1+z)^{1/3}f^{4/3}\right)^{3/2}. 
\label{l_of_hzf}
\eeq

Cusp formation is expected to occur ${\cal O}(1)$ times in an
oscillation period, which is characterized by parameter $c$.  The
value of $c$ can be made to correspond to the emission efficiency
$\Gamma$ \cite{Binetruy:2012ze}, and we use
$c=2^{1/3}\Gamma/(3\pi^2)\eqsim 2.13$.  Then the number of GWs coming
to the Earth per unit time, emitted at redshift between $z$ and $z+dz$
by loops formed between $t_i$ and $t_i+dt_i$, is given using the
loop-number density obtained in the previous subsection as
\beq
\frac{dR}{dzdt_i}dzdt_i=\frac{1}{4}\theta_m (f,z,l)^2\frac{2c}{ (1+z)l (t (z),t_i)}
\frac{dn}{dt_i} (t (z),t_i)dt_i\cdot\frac{dV}{dz}dz\cdot\Theta (2-\theta_m (f,z,l)), \label{dR_dtdti}
\eeq
where $\theta_m$ is the beaming angle of the GW burst and given by
\beq
\theta_m (f,z,l)= ( (1+z)fl)^{-1/3}, \label{thetam}
\eeq
and
\beq
\frac{dV}{dz} (z)=\frac{4\pi a^2 (z)r^2 (z)}{H (z) (1+z)}.
\eeq
The factor $\frac{1}{4}\theta_m (f,z,l)^2$ reflects the beaming of the
GW bursts, and the Heviside step function $\Theta$ reflects the
low-frequency cutoff of $f l \lesssim 2$ at the emission
\cite{Binetruy:2012ze}.  Using Eqs. (\ref{length}) and (\ref{strain}),
we can rewrite Eq. (\ref{dR_dtdti}) to express the number of GWs
coming per unit time which were emitted at redshift $z$ and which have
frequency $f$ and amplitude $\tilde{h}$ at the present time:
\beq
\frac{dR}{dzd\tilde{h}} (f,\tilde{h},z)=\frac{3}{4}\theta^2_m (f,z,l)\frac{c}{ (1+z)\tilde{h}}\frac{1}{\gamma(t_i)^2 \alpha t^4_i}
\frac{1}{\alpha+\Gamma G\mu}\left (\frac{a (t_i)}{a (t)}\right)^3 \frac{dV}{dz}
\Theta  (2-\theta_m (f,z,l)).
\label{dR_dzdh}
\eeq
By integrating the rate along the redshift, we get the total arrival
rate of GWs today,
\beq
\frac{dR}{d\tilde{h}}(f,\tilde{h})=\int^{\infty}_0 dz \frac{dR}{dzd\tilde{h}}(f,\tilde{h},z).
\label{dR_dh}
\eeq

\subsection{GW background}
The amplitude of the GW background is characterized by the
dimensionless parameter $\Omega_{\rm GW} (f)\equiv (d\rho_{\rm
  GW}/d\ln f)/\rho_{cr}$, where $\rho_{\rm GW}$ is the energy density
of the GWs and $\rho_{cr}$ is the critical density of the Universe.
By summing up all the bursts, $\Omega_{\rm GW}$ is given by
\beq
\Omega_{\rm GW} (f)=\frac{2\pi^2}{3H_0^2}f^3\int^{\tilde{h}_*}_0d\tilde{h}\tilde{h}^2\frac{dR}{d\tilde{h}}(f,\tilde{h}),
\label{OmegaGW}
\eeq
where $H_0=100 h$ km/s/Mpc is the Hubble parameter
at the present time.  The condition of forming a stochastic GW
background is commonly implemented by introducing the upper limit of
the integration $\tilde{h}_*$, which satisfies \cite{Siemens:2006yp}
\beq
\int^{\infty}_{\tilde{h}_*} d\tilde{h}\frac{dR}{d\tilde{h}}=f.
\eeq
By setting this, the integration of the GW amplitude is carried out to
include only bursts with small $\tilde{h}$ which come to the observer
with a time interval shorter than $1/f$.  Such bursts overlap each
other and are considered to be unresolved as a single burst, while
rare bursts with large $\tilde{h}$ are observed individually.

\begin{figure}[!t]
  \begin{center}
  \includegraphics[width=0.48\textwidth]{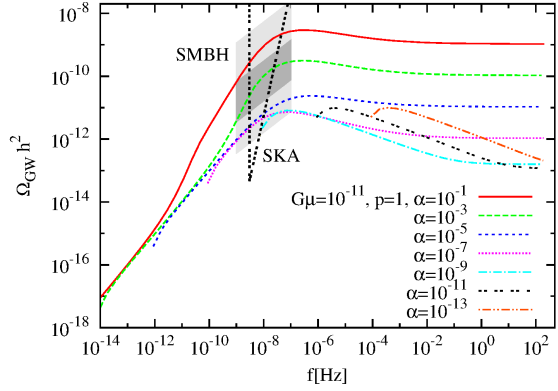}
  \end{center}
  \caption{ The GW spectra originating from cusps on cosmic string
    loops are plotted in terms of $\Omega_{\rm GW}h^2$ by changing the
    values of the initial loop size $\alpha$.  The cosmic string
    tension is fixed at $G\mu=10^{-11}$, and the reconnection
    probability is $p=1$.  The sensitivity of the SKA is shown by the
    black dotted line.  The shaded areas represent the 68$\%$ (dark
    gray) and 99.7$\%$ (light gray) confidence intervals of the
    theoretically predicted GW amplitude from SMBH binaries according
    to Ref. \cite{Sesana:2012ak}.}
\label{fig:OGW}
\end{figure}

\begin{figure*}[!t]
 \begin{minipage}{0.48\hsize}
  \begin{center}
    \includegraphics[width=100mm]{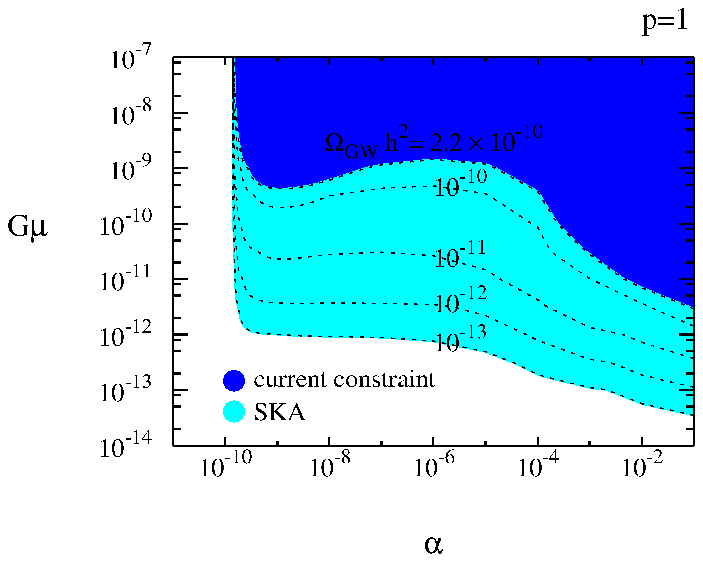}
  \end{center}
 \end{minipage}
 \begin{minipage}{0.48\hsize}
 \begin{center}
    \includegraphics[width=100mm]{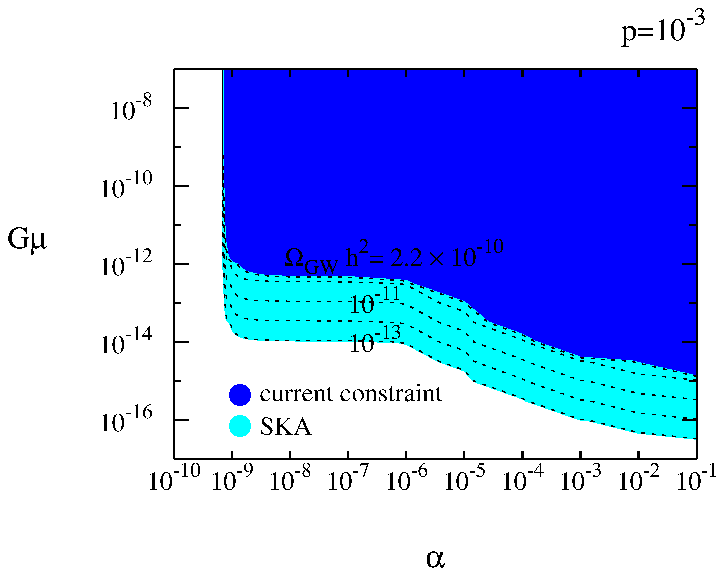}
 \end{center}
 \end{minipage}
 \caption{ Parameter space accessible by SKA (light blue) for $p=1$ and
   $p=10^{-3}$.  The blue region is the space which is already
   excluded by the current pulsar constraints.  The black dotted lines
   correspond to contour lines of $\Omega_{\rm
     GW}h^2=10^{-13},10^{-12},10^{-11},10^{-10}$, and $2.2\times10^{-10}$ from
   bottom to top. }
\label{fig:dRdlnh_ex}
\end{figure*}

Using the formalisms described above, we show $\Omega_{\rm GW}h^2$ as
a function of frequency for various values of the initial loop size
parameter $\alpha$ in Fig. \ref{fig:OGW}.  Figure \ref{fig:dRdlnh_ex}
shows the parameter regions excluded by the current strongest pulsar
experiments, as well as regions which can be probed by SKA.  For the
excluded regions, we use the most stringent limit of the
scale-invariant spectrum obtained by NANOGrav
\cite{Arzoumanian:2015liz}, $\Omega_{\rm GW}h^2 < 2.2\times 10^{-10}$.
And for the SKA-accessible region, we show the region where
$\Omega_{\rm GW}h^2 > 10^{-13}$.

\section{Anisotropies in the GW background}
\label{sec:anisotropy}

\subsection{Method}
\label{animethod}

In order to estimate the anisotropies in the GW background, we
simulate the GW background by distributing GW sources on the sky map.
First, we compute the GW rate as a function of the amplitude
$\tilde{h}$.  The GW burst rate for a given amplitude $\tilde{h}$ and
for a fixed frequency can be calculated by Eq. (\ref{dR_dh}).  Then we
distribute GW bursts on the sky according to the obtained burst rate.
The position is randomly assigned to each source.  We generate 100
realizations of the sky map and calculate the average and variance of
the anisotropy level by following the method described in
Ref. \cite{Taylor:2013esa}.  The formalism is developed in
Ref. \cite{Mingarelli:2013dsa}.  First, we decompose the energy
density of GWs $\rho(\hat\Omega)\propto \tilde{h}^2$ in terms of the
spherical harmonic functions as
\beq
\rho(\hat\Omega) = \sum^{\infty}_{\ell=0}\sum^{\ell}_{m=-\ell}c_{\ell m}Y_{\ell m}(\hat\Omega),
\label{rho}
\eeq
where $\hat{\Omega}$ represents the propagation direction of the GW.
For point sources, the anisotropy coefficients can be calculated by
\beq
c_{\ell m} = \sum_{i=1}^N\; \rho_i Y_{\ell m}(\hat\Omega_i),
\label{clm}
\eeq
where $\rho_i$ describes the GW energy density of each source.  Using
the definition of the angular power spectrum $C_\ell = \sum_m
|c_{\ell m}|^2/(2\ell+1)$, we calculate the anisotropic power normalized by the
monopole component $C_\ell/C_0$.

The above method is identical to decomposing $\Omega_{\rm GW}$ as
\cite{Mingarelli:2013dsa}
\beq
\Omega_{\rm GW}(f) = \frac{2 \pi^2}{3 H_0^2} f^3 \cdot 4 H(f)  
\int d\hat{\Omega}\, P(\hat{\Omega})\,,
\eeq
with
\beq
P(\hat\Omega) = \sum^{\infty}_{\ell=0}\sum^{\ell}_{m=-\ell}\tilde{c}_{\ell m}Y_{\ell m}(\hat\Omega).
\eeq
In this case, the anisotropy coefficients are calculated by
\beq
\tilde{c}_{\ell m}=\int d\hat{\Omega}\, P(\hat{\Omega})Y_{\ell m}(\hat\Omega).
\eeq
Here, $H(f)$ corresponds to the spectral power of the isotropic
(monopole) component, and $P(\hat{\Omega})$ describes the angular
distribution of the anisotropic components, which is related to
Eq. (\ref{OmegaGW}) by $\int d\tilde{h}
\tilde{h}^2\frac{dR}{d\tilde{h}}=\tilde{h}_{\rm
  rms}^2(f,\hat\Omega)\equiv 4H(f)\int d\hat\Omega P(\hat\Omega)$.
Note that $\tilde{c}_{\ell m}$ is normalized by
$\tilde{c}_{00}=\sqrt{4\pi}$, which gives $\int d\hat{\Omega}\,
P(\hat{\Omega}) = 4 \pi$, while $c_{\ell m}$ in Eq. (\ref{rho})
includes coefficients of the GW power.  This does not make a
difference in the results as long as we use the normalized power
spectrum $C_\ell/C_0$.

\begin{figure*}[htbp]
 \begin{minipage}{0.3\hsize}
  \begin{center}
   \includegraphics[width=52mm]{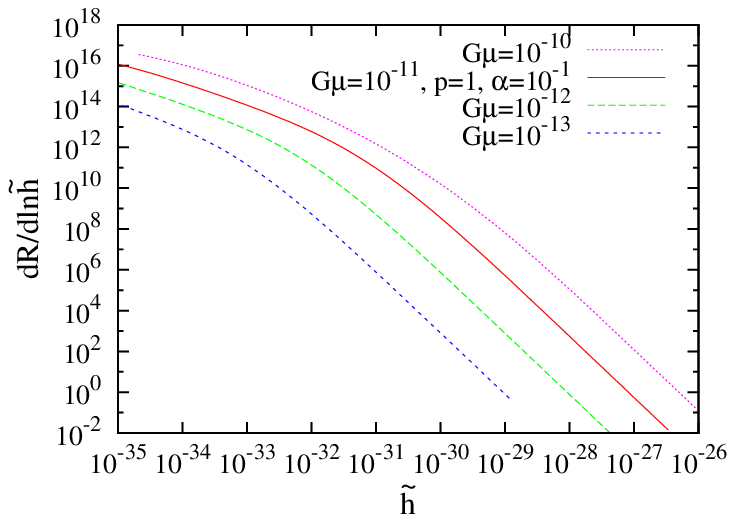}
  \end{center}
 \end{minipage}
 \begin{minipage}{0.3\hsize}
 \begin{center}
   \includegraphics[width=52mm]{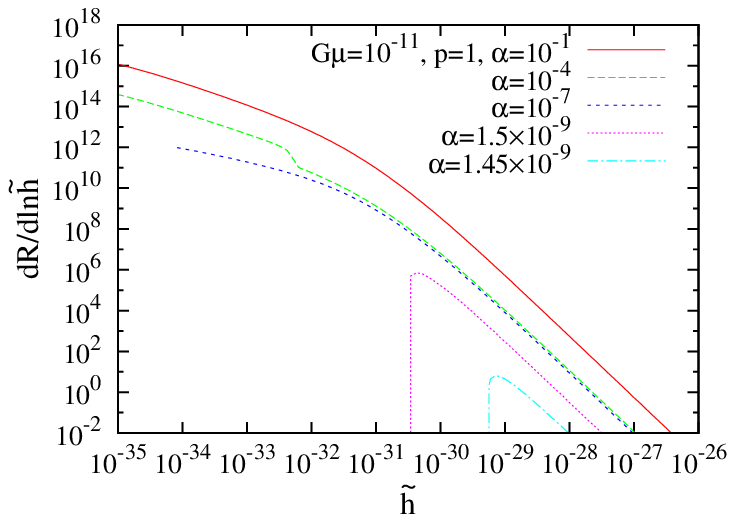}
 \end{center}
 \end{minipage}
 \begin{minipage}{0.3\hsize}
 \begin{center}
   \includegraphics[width=52mm]{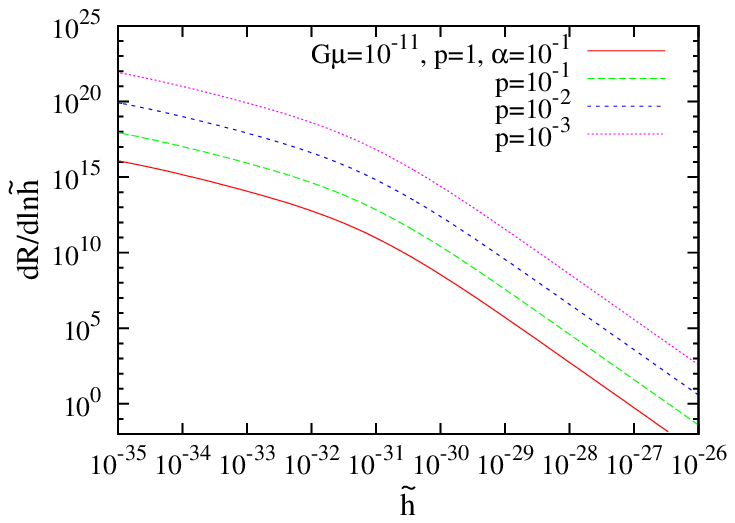}
 \end{center}
 \end{minipage}
 \caption{The expected number of GW bursts per logarithmic strain,
   $dR/d{\ln}\tilde{h}$, for the fixed frequency $f=1/10{\rm years}=3.17\times
   10^{-9}{\rm Hz}$.  From the left to the right, we show parameter
   dependencies of $G\mu$, $\alpha$, and $p$.  The vertical axis is
   expected numbers of GW bursts per 10 years.  }
  \label{fig:rate}
\bigskip
 \begin{minipage}{0.3\hsize}
  \begin{center}
   \includegraphics[width=52mm]{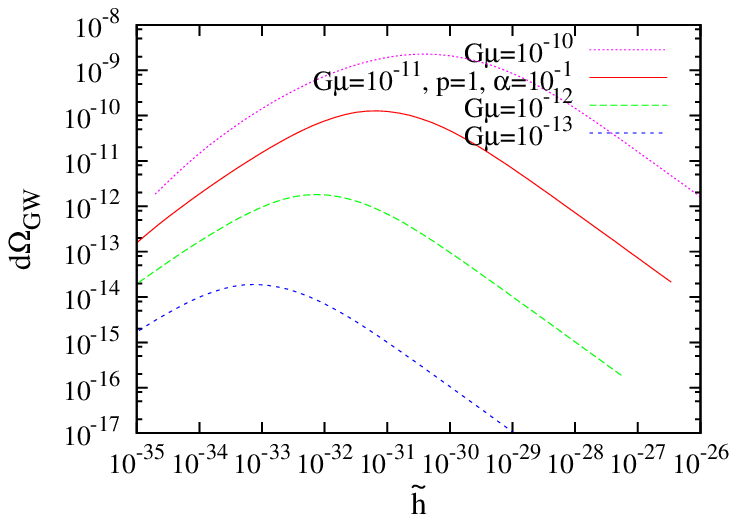}
  \end{center}
 \end{minipage}
 \begin{minipage}{0.3\hsize}
 \begin{center}
  \includegraphics[width=52mm]{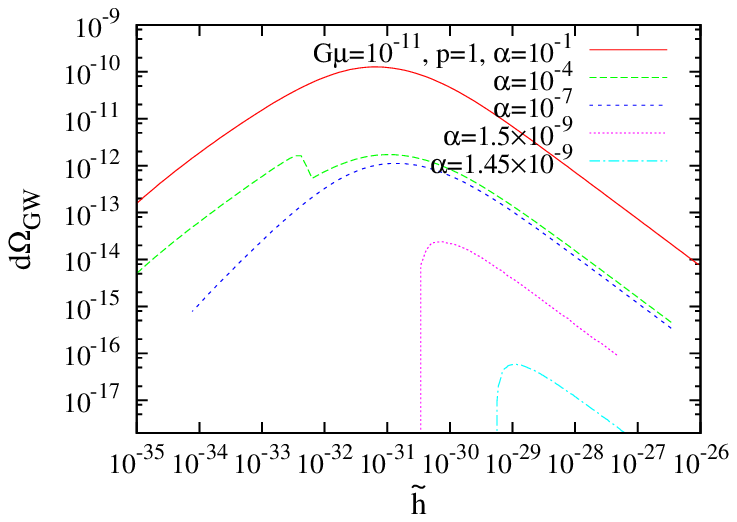}
 \end{center}
 \end{minipage}
 \begin{minipage}{0.3\hsize}
 \begin{center}
  \includegraphics[width=52mm]{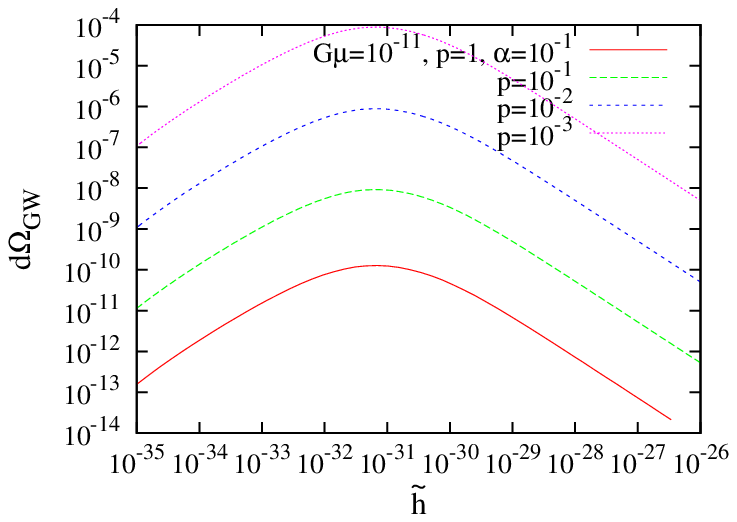}
 \end{center}
 \end{minipage}
 \caption{ The contribution to the integral of Eq. (\ref{OmegaGW}) for
   each logarithmic strain bin with the same parameter set as in
   Fig. \ref{fig:rate}. }
  \label{fig:dOmegaGW}
\bigskip
 \begin{minipage}{0.3\hsize}
  \begin{center}
   \includegraphics[width=52mm]{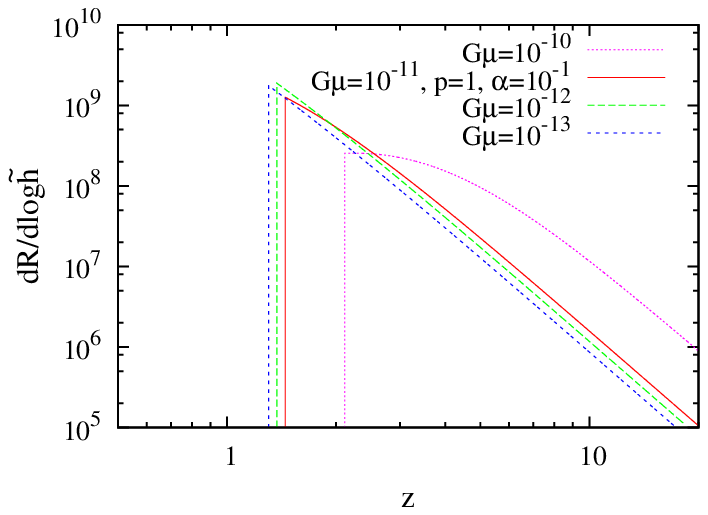}
  \end{center}
 \end{minipage}
 \begin{minipage}{0.3\hsize}
 \begin{center}
  \includegraphics[width=52mm]{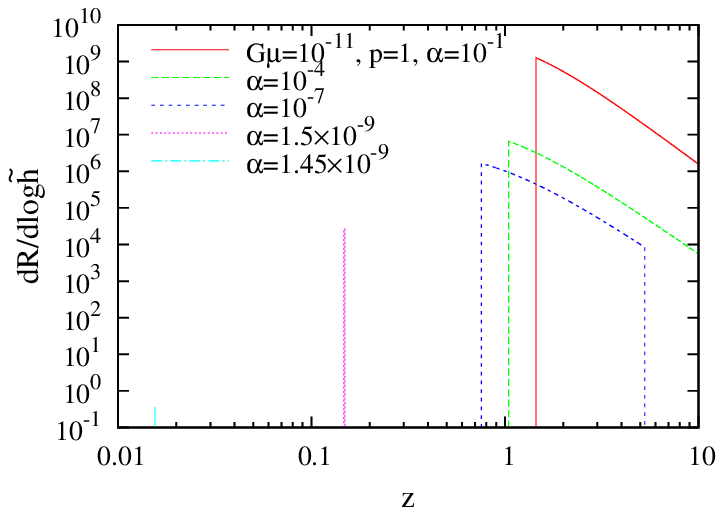}
 \end{center}
 \end{minipage}
 \begin{minipage}{0.3\hsize}
 \begin{center}
  \includegraphics[width=52mm]{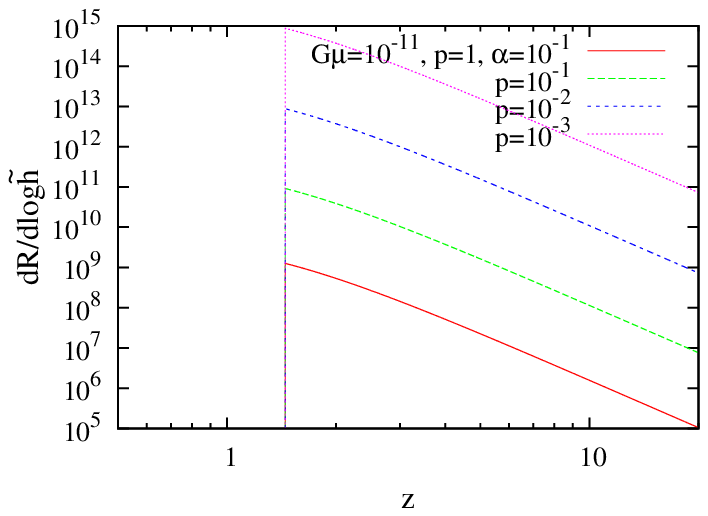}
 \end{center}
 \end{minipage}
 \caption{ The rate shown in terms of the redshift with the same
   parameter set as in Fig. \ref{fig:rate}.  We fix the strain
   amplitude to be the one which gives the biggest contribution to the
   value of $\Omega_{\rm GW}$ (the value of $\tilde{h}$ at the local
   maximum in Fig. \ref{fig:dOmegaGW}). }
  \label{fig:zdistribution}
\end{figure*}

To calculate the anisotropy coefficients, we use Eq. (\ref{clm}) and
sum up all the energy density sources in the simulated sky map.  Note
that the inclusion of all the sources means that we do not apply the
upper limit $\tilde{h}_*$ in Eq. (\ref{OmegaGW}).  This treatment may not be
proper when strong bursts are identified and removed from the
background data.  However, since the observation time of pulsar timing
is comparable to the time scale of the GW period $T_{\rm obs}\sim
1/f$, it would be difficult to resolve a single burst, and rare bursts
would not be distinguished from the GW background.  In fact, the
inclusion of $\tilde{h}_*$ affects the spectrum shape only at higher
frequencies and is important for direct detection by interferometers.

\subsection{Rate of GW bursts}

Using Eq. (\ref{dR_dh}), we can predict the rate of the GW bursts for
a given frequency and given parameter values.  Let us first see the
parameter dependence of the rate of the GW bursts.  Figure
\ref{fig:rate} shows the GW rate at $f=1/10{\rm years}=3.17\times
10^{-9}{\rm Hz}$ per 10 years for different values of the cosmic
string parameters -- the tension $G\mu$, the initial loop size
$\alpha$, and the reconnection probability $p$.  The contribution to
$\Omega_{\rm GW}$ is determined by $\tilde{h}^2\frac{dR}{d\tilde{h}}$,
as seen in Eq. (\ref{OmegaGW}).  In Fig. \ref{fig:dOmegaGW}, we plot
the derivative contribution to $\Omega_{\rm GW}$ in terms of
$\tilde{h}$ to find the strain amplitude which mainly composes the GW
background.  In Fig. \ref{fig:zdistribution}, we show the redshift
evolution of the rate using Eq. (\ref{dR_dzdh}).  To calculate the
redshift dependence of the rate $\frac{dR}{d\tilde{h} dz}$, we need to
fix the value of $\tilde{h}$.  We choose $\tilde{h}$ to be the local
maximum of each line in Fig. \ref{fig:dOmegaGW} -- that is, the value of
$\tilde{h}$ which contributes the most to $\Omega_{\rm GW}$.

The small jumps seen in the middle panels of Figs. \ref{fig:rate} and
\ref{fig:dOmegaGW} for $\alpha=10^{-4}$ are an artifact of the sudden
transition of $\gamma$ from the radiation-dominated to the
matter-dominated era as provided in Eq. (\ref{gamma_rm}).  For this
low frequency $f=3.17\times 10^{-9}{\rm Hz}$, all GWs are emitted
after radiation-matter equality, but the loops are formed in the
radiation-dominated era when they have a long lifetime
$\alpha=10^{-1}$, while loops are formed in the matter-dominated era
when they have a short lifetime $\alpha\leq 10^{-7}$.  The
intermediate case is $\alpha=10^{-4}$, where GWs with small
$\tilde{h}$ are emitted from loops formed during the
radiation-dominated era, while a large $\tilde{h}$ corresponds to
loops formed in the matter-dominated era.

To achieve the detectable large anisotropy of $\sim\mathcal{O}(10\%)$
in the GW background, a small number of strong bursts should
contribute the GW background comparable to the overall amplitude.  As
seen from Figs. \ref{fig:dOmegaGW} and \ref{fig:zdistribution}, in
most of the cases, the dominant component of the GW background is the
numerous small bursts coming from high redshifts, where we cannot
expect large anisotropy.  The interesting case is found when the
initial loop size $\alpha$ is small (see the middle panels).  The
lifetime of the cosmic string loop is given by the initial energy of
the loop divided by the rate of the energy release by GW emissions,
$\tau\sim \alpha t_i/(\Gamma G\mu)$.  Thus, when $\alpha \ll \Gamma
G\mu$, loops decay within a Hubble time after their formation.  GWs
emitted from such short-lived loops have a typical frequency which
corresponds to their loop size $(1+z_i)f\sim 2/(\alpha t_i)$.  Because
of this, the GW background of a given frequency consists of GW bursts
from a specific redshift, as seen from Fig. \ref{fig:zdistribution}.
For a fixed frequency $f=[\alpha t_i(1+z_i)]^{-1}$, $t_i$ should
increase for smaller $\alpha$.  This leads to a lower number density
of loops, since $a(t_i)^3/t_i^{4}$ in Eq. (\ref{dn_dt2}) is a
decreasing function with respect to $t_i$.  Therefore, as seen in
Fig. \ref{fig:rate}, we find the case where the GW background consists
of a small number of bursts for a specific range of $\alpha$.  In this
case, we can expect a large anisotropy as shown in the next section.

\subsection{Results and discussions}

\begin{figure}[!t]
  \begin{center}
  \includegraphics[width=0.48\textwidth]{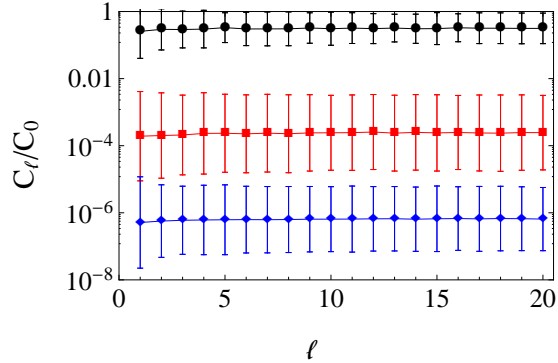}
  \end{center}
  \caption{ The anisotropy power $C_\ell/C_0$ for different values of
    the initial loop size: $\alpha=1.45\times 10^{-9}$ (black),
    $1.5\times 10^{-9}$ (red), and $10^{-4}$ (blue).  We set
    $G\mu=10^{-11}$ and $p=1$ and assume observation at $f=1/10{\rm
      year}=3.17\times 10^{-9}{\rm Hz}$. }
\label{fig:ani_alpha}
\end{figure}

Finally, we estimate the anisotropy level using the method described in
Sec. \ref{animethod}.  In Fig. \ref{fig:ani_alpha}, we show anisotropy
power $C_\ell/C_0$ up to the multipole $\ell=20$ for
$\alpha=1.45\times 10^{-9}$, $1.5\times 10^{-9}$, and $10^{-4}$.  The
other parameters are fixed at $G\mu=10^{-11}$ and $p=1$.  The central
point is the mean value of $100$ realizations, and the error bars
represent the $2\sigma$ variances.  Since we find the distribution is
near the log-normal Gaussian distribution, we calculate the mean
values and variances for logarithmic values of $C_\ell/C_0$.  We see
that the anisotropy becomes large even to a level that could be
detected by SKA $\sim \cal{O}$(0.1) in the case of $\alpha=1.45\times
10^{-9}$, but it decreases quickly when $\alpha$ is reduced to
$\alpha=1.5\times 10^{-9}$.

\begin{figure}[!t]
  \begin{center}
  \includegraphics[width=0.48\textwidth]{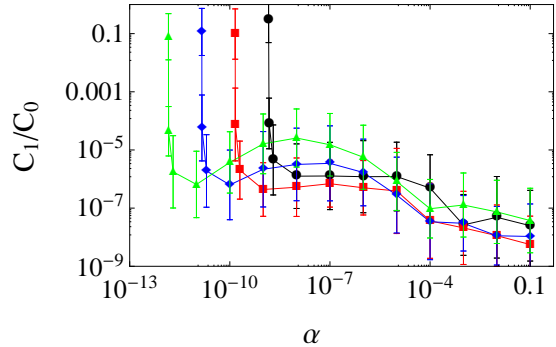}
  \end{center}
  \caption{ The anisotropy power $C_\ell/C_0$ of the dipole moment
    shown as a function of $\alpha$.  We assume $G\mu=10^{-11}$ and
    $p=1$.  The different colors describe different observation
    frequencies; $f=1/10{\rm year}=3.17\times 10^{-9}{\rm Hz}$
    (black), $3.17\times 10^{-8}{\rm Hz}$ (red), $3.17\times
    10^{-7}{\rm Hz}$ (blue), and $3.17\times 10^{-6}{\rm Hz}$
    (green).}
\label{fig:ani_dipole}
\end{figure}

Since the anisotropy is the same level for all the multipoles as seen
from Fig. \ref{fig:ani_alpha}, let us focus on the dipole moment from
now on.  In Fig. \ref{fig:ani_dipole}, we plot the dipole power
$C_1/C_0$ as a function of $\alpha$.  The different lines correspond
to different observation frequency bands.  As $\alpha$ decreases, the
anisotropy power suddenly increases because of the decrease in the
number density of the loops.  We do not have points in the region
where $\alpha$ is smaller than the peak point, since GWs are not
generated in pulsar timing frequencies due to the low-frequency cutoff
of the GW emission, $f\lesssim 2/l$.  In other words, detecting the
strong anisotropy power corresponds to observing the edge of the
spectrum at the lowest frequency.  As in Fig. \ref{fig:OGW}, the
amplitude of the spectrum decreases towards the edge because of the
limited number of the loops.  The edge shifts to the higher frequency
for smaller $\alpha$, and we get no GW power when the spectrum goes out
of the sensitivity range of the pulsar timing experiments.

A large anisotropy is achieved only when the GW background consists of
bursts from new loops near us without having any contribution from old
loops, which are numerous and reduce the anisotropy level.  Such
conditions are satisfied around the low-frequency cutoff $f\sim
2/(\alpha t_0)$, which is the typical frequency of bursts from
recently formed loops, and old loops cannot contribute to this
frequency, since their size must be smaller.  Therefore, the position
of the anisotropy peak depends on the value of $\alpha$ as well as the
observation frequency $f$.  For the typical pulsar timing observation
frequency $f\sim 1/10$ years, the peak arises at $\alpha\sim
2/(ft_0) \sim 10^{-9}$.  The peak position moves when we change the
observation frequency, as seen in Fig. \ref{fig:ani_dipole}.

\begin{figure*}[!t]
 \begin{minipage}{0.48\hsize}
  \begin{center}
  \includegraphics[width=80mm]{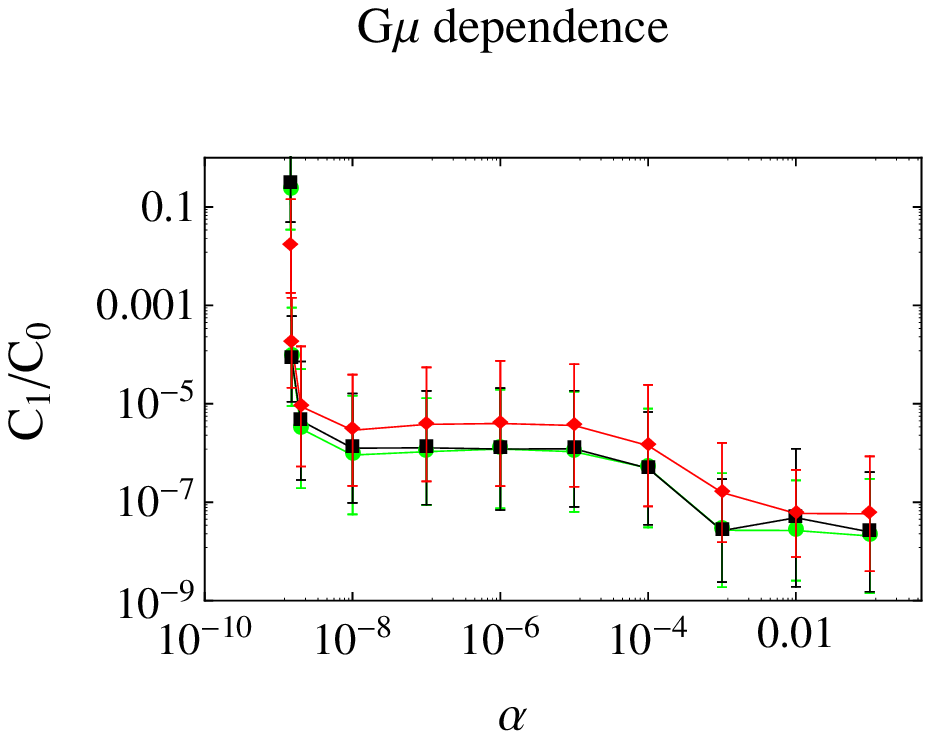}
  \end{center}
 \end{minipage}
 \begin{minipage}{0.48\hsize}
  \begin{center}
    \includegraphics[width=80mm]{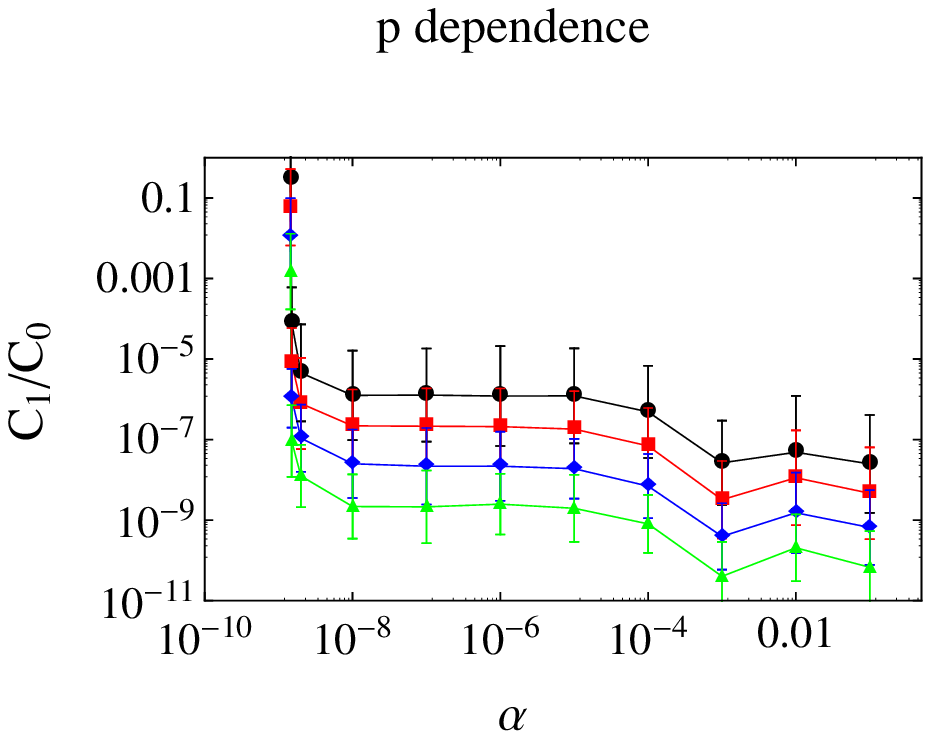}
  \end{center}
 \end{minipage}
 \caption{ The anisotropy power $C_\ell/C_0$ of the dipole moment
   shown as a function of $\alpha$.  The observation frequency is
   assumed to be $f=1/10{\rm year}=3.17\times 10^{-9}{\rm Hz}$.  In
   the left panel, we change the value of tension $G\mu$ by fixing
   $p=1$; $G\mu=10^{-10}$ (red), $10^{-11}$ (black), and $10^{-12}$
   (green).  In the right panel, we change the value of reconnection
   probability by fixing $G\mu=10^{-11}$; $p=1$ (black), $10^{-1}$
   (red), $10^{-2}$ (blue), and $10^{-3}$ (green).}
\label{fig:ani_dipole_gmu_p}
\end{figure*}

In Fig. \ref{fig:ani_dipole_gmu_p}, we show the dependence of our
result on other parameters, such as tension $G\mu$ and reconnection
probability $p$.  We find that the amplitude of the anisotropies
changes depending on the parameters, but the position of the peak does
not change.  The anisotropy amplitude is determined by the number of
bursts which form the GW background.  We see that the anisotropy is
reduced for smaller value of $p$, because small $p$ simply increases
the overall amplitude of the number distribution.  The value of $G\mu$
changes the number distribution as well as the value of $\tilde{h}$,
which gives the main contribution to $\Omega_{\rm GW}$, as seen in
Figs. \ref{fig:rate} and \ref{fig:dOmegaGW}.  The combination of the
two effects turns out to be a small decrease in the burst number for
larger $G\mu$, which is the reason we see the anisotropy gets slightly
larger for $G\mu=10^{-10}$.  In contrast to the anisotropy amplitude,
we find that the peak position does not move for any change of $G\mu$
and $p$.  This is because the condition to have a large anisotropy is
$f\sim 2/(\alpha t_0)$, which does not depend on the values of $G\mu$
and $p$.

\begin{figure}[!t]
  \begin{center}
  \includegraphics[width=0.48\textwidth]{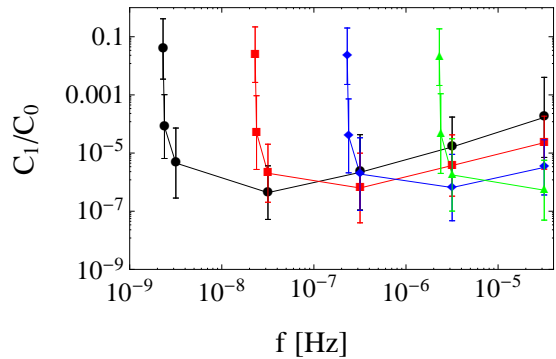}
  \end{center}
  \caption{ The anisotropy power $C_\ell/C_0$ of the dipole moment
    shown as a function of the observation frequency.  The different
    colors describe different values of the initial loop size;
    $\alpha=10^{-9}$ (black), $10^{-10}$ (red), $10^{-11}$ (blue), and
    $10^{-12}$ (green).  The other parameters are set as
    $G\mu=10^{-11}$ and $p=1$.}
\label{fig:ani_dipole_f}
\end{figure}

As shown in Fig. \ref{fig:ani_dipole}, one can expect a large
anisotropy for different values of $\alpha$ by changing the
observation frequency.  In Fig. \ref{fig:ani_dipole_f}, we plot the
result as a function of the frequency for different values of
$\alpha$.  When one analyzes a specific frequency bin, the anisotropy
is small for the most of the values of $\alpha$, but one can expect a
large anisotropy for a specific value of $\alpha$.  This indicates
that one can, in principle, test the value of $\alpha$ by checking the
anisotropy power at different frequency bands of GWs.  A typical GW
frequency of the pulsar timing array is $f\sim 1/10{\rm
  years}=3.17\times 10^{-9}{\rm Hz}$, while we would be able to
analyse higher-frequency bands possibly up to $10^{-6}$, which is
limited by the monitoring time interval of pulsar observation.  Thus,
by analyzing frequency bands of $10^{-9} \lesssim f \lesssim 10^{-6}$,
we may be able to probe the range of $10^{-11} \lesssim \alpha
\lesssim 10^{-9}$.

As for the current constraint, the European Pulsar Timing Array
\cite{Taylor:2015udp} has placed limits on the multipole components of
the GW amplitude for $\ell>0$ is less than $40\%$ of the isotropic
component at $2-90\times 10^{-9}$Hz.  Thus, since the result does not
depend on the value of $G\mu$, we can say that $5\times 10^{-11} <
\alpha < 2\times 10^{-9}$ is excluded for $G\mu \gtrsim 1.3\times
10^{-7}$ with $p=1$ \cite{Lentati:2015qwp}. (Note that we can only
exclude the region where the tension is constrained by the multipole
component, since the anisotropy level is defined as the ratio to the
monopole $C_\ell/C_0$. )  This corresponds to excluding the left edge
of the blue region in Fig. \ref{fig:dRdlnh_ex}.

One may express concern about the assumption of the one-scale model
which enforces the uniform initial loop size, while the initial loop
size $\alpha$ could be distributed around a typical value.  The shape
of the GW rate would change when one takes into account the
distribution of $\alpha$, but the total number of bursts is important,
rather than the shape, for the estimation of the anisotropy level.
Thus, we would still find the case where the background consists of a
few sources and has a large anisotropy.  However, it would be
difficult to extract information on the distribution of $\alpha$ from
the anisotropy power unless we imposed a simple model for the
distribution.  This argument would apply also to the case for
different models of the cosmic string network.  Since the number of
bursts is the only key factor for anisotropy, the existence of the
strong anisotropy would be universal for most cosmic string models.

Another point we may have to take into account is the relativistic
nature of small loops.  References
\cite{Polchinski:2006ee,Blanco-Pillado:2013qja} pointed out that small
loops are created with ultrarelativistic speeds.  This would be the
case for our targeting range of loop size $10^{-11} \lesssim \alpha
\lesssim 10^{-9}$.  The relativistic motion of loops would give rise
to a blueshift of the GW frequency, which would shift the peak
position to large $\alpha$.  It would also reduce the beaming angle,
and therefore the amplitude of the GW burst would increase while the
event rate decreased.  This would also affect the peak position, but
since these two effects compensate each other, further study is
necessary to quantitatively estimate the impact of the relativistic
speed of loops on the peak position.

Finally, let us comment on the possibility of testing of the
anisotropic GW background by laser-interferometer experiments such as
Advanced LIGO \cite{Abadie:2011rr,Thrane:2009fp}.  We would expect the
same result with a different peak position for LIGO, which is
determined by $f_{\rm LIGO}\sim 2/(\alpha t_0)\sim 100$Hz.  An
anisotropic background may be rephrased as a noncontinuous background,
which is known to be testable by examining the non-Gaussianity of the
data.  Reference \cite{Regimbau:2011bm} estimates the possibility of
detecting such a noncontinuous popcorn-like background from cosmic
strings with ground-based laser interferometers.  It also estimates
the case of pulsar timing frequency $f=10^{-8}$Hz and shows that the
popcorn feature arises at $\epsilon\sim 1$ for $G\mu=10^{-11}$, where
$\alpha=\epsilon\Gamma G\mu$.  This is consistent with our result.

\section{Conclusion}
\label{sec:conclusion}

We have investigated the possibility of having an anisotropic GW
background originating from cosmic string loops.  The anisotropy turns
out to be too small to be detected by pulsar timing experiments in
most parameter spaces, while it becomes large for a specific value of
the initial loop size $\alpha$.  The large anisotropy is found when a
small number of GW bursts contributes to the observed GW background.
We found that the parameter space of $\alpha$ yielding a strong
anisotropy is very narrow when we analyze the data at a fixed
frequency, but one can access $10^{-11} \lesssim \alpha \lesssim
10^{-9}$ by analyzing in different frequency bands.

To have the detectable anisotropy of ${\cal O}(10\%)$, we need the
bright outlier source which dominates the overall components of the GW
background, as is the same for the SMBH-binary background
\cite{Taylor:2013esa}.  The existence of such an outlier may not be
extremely rare in the case of SMBH binaries, as some theoretical
studies indicate that sources at redshift $0.1-1$ could be bright
enough to be individually resolved \cite{Sesana:2008xk}.  On the other
hand, in the case of the GW background from cosmic strings, distant
sources are typically dominant in number.  We have found that the
nearest sources become dominant and produce an anisotropic GW
background only for the specific parameter choice where the initial
loop size $\alpha$ is very small.  However, since the properties of
the cosmic string network are not understood very well yet, the
anisotropy test is still useful for exploring a new parameter space
and helps us to understand its distribution.

So far, only the amplitude of the GW background has been used to place
constraints on the cosmic string parameters, and the parameter
degeneracies cannot be removed only by the information of the amplitude.
Although the estimation of the sensitivity for the anisotropy test in
multiple frequency bins is beyond our work, anisotropy may provide a
new and independent opportunity to constrain the value of $\alpha$ and
help to test the string network models.

\section*{Acknowledgements}
S.K. is supported by the Career Development Project for Researchers of
Allied Universities. K.T. is supported by Grand-in-Aid from the Ministry
of Education, Culture, Sports, and Science and Technology (MEXT) of
Japan, No. 24340048, No. 26610048, and No. 15H05896.

{}

\end{document}